      \documentstyle[12pt,graphicx]{article}                
\begin{document}
\baselineskip 18pt
\def\today{\ifcase\month\or
 January\or February\or March\or April\or May\or June\or
 July\or August\or September\or October\or November\or December\fi
 \space\number\day, \number\year}
\def\thebibliography#1{\section*{References\markboth
 {References}{References}}\list
 {[\arabic{enumi}]}{\settowidth\labelwidth{[#1]}
 \leftmargin\labelwidth
 \advance\leftmargin\labelsep
 \usecounter{enumi}}
 \def\newblock{\hskip .11em plus .33em minus .07em}
 \sloppy
 \sfcode`\.=1000\relax}
\let\endthebibliography=\endlist
\def\lsim{\ ^<\llap{$_\sim$}\ }
\def\gsim{\ ^>\llap{$_\sim$}\ }
\def\r2{\sqrt 2}
\def\beq{\begin{equation}}
\def\eeq{\end{equation}}
\def\beqn{\begin{eqnarray}}
\def\eeqn{\end{eqnarray}}
\def\rmuu{\gamma^{\mu}}
\def\rmud{\gamma_{\mu}}
\def\PL{{1-\gamma_5\over 2}}
\def\PR{{1+\gamma_5\over 2}}
\def\sinW2{\sin^2\theta_W}
\def\AEM{\alpha_{EM}}
\def\mul{M_{\tilde{u} L}^2}
\def\mur{M_{\tilde{u} R}^2}
\def\mdl{M_{\tilde{d} L}^2}
\def\mdr{M_{\tilde{d} R}^2}
\def\mz2{M_{z}^2}
\def\c2b{\cos 2\beta}
\def\au{A_u}
\def\ad{A_d}
\def\cob{\cot \beta}
\def\v#1{v_#1}
\def\tb{\tan\beta}
\def\epem{$e^+e^-$}
\def\KK{$K^0$-$\bar{K^0}$}
\def\wi{\omega_i}
\def\xj{\chi_j}
\def\Wmu{W_\mu}
\def\Wnu{W_\nu}
\def\m#1{{\tilde m}_#1}
\def\mH{m_H}
\def\mw#1{{\tilde m}_{\omega #1}}
\def\mx#1{{\tilde m}_{\chi^{0}_#1}}
\def\mc#1{{\tilde m}_{\chi^{+}_#1}}
\def\mwi{{\tilde m}_{\omega i}}
\def\mxi{{\tilde m}_{\chi^{0}_i}}
\def\mci{{\tilde m}_{\chi^{+}_i}}
\def\mz{M_z}
\def\sw{\sin\theta_W}
\def\cw{\cos\theta_W}
\def\cb{\cos\beta}
\def\sb{\sin\beta}
\def\rwi{r_{\omega i}}
\def\rxj{r_{\chi j}}
\def\rfp{r_f'}
\def\Kik{K_{ik}}
\def\Fq2{F_{2}(q^2)}
\def\tw{\tan\theta_W}
\def\sec2w{sec^2\theta_W}

\begin{titlepage}

\  \
\vskip 0.5 true cm
\begin{center}
{\large {\bf CP Violation and Dark Matter }}\\

\vskip 0.5 true cm
\vspace{2cm}
\renewcommand{\thefootnote}
{\fnsymbol{footnote}}
 Utpal Chattopadhyay$^a$, Tarek Ibrahim$^b$ and Pran Nath$^{c}$  
\vskip 0.5 true cm
\end{center}
\noindent
{a. Tata Institute of Fundamental Research, Homi Bhabha Road,
 Mumbai 400 005, India }\\
{b. Department of  Physics, Faculty of Science,
University of Alexandria,}\\
{ Alexandria, Egypt}\\ 
{c. Department of Physics, Northeastern University,
Boston, MA 02115-5000, USA } \\
\vskip 1.0 true cm

\centerline{\bf Abstract}
\medskip
A brief review is  given of the effects of CP violation on 
the direct detection of neutralinos in dark matter detectors.
We first summarize the current developments using the cancellation 
mechanism which allows for the existence of large CP violating
phases consistent with experimental limits on the electron and 
on the neutron electric dipole moments in a broad class of SUSY,
string and D brane models. We then discuss their 
effects on the scattering of neutralinos from quarks and 
on the event rates. It is found that while CP effects on the event
rates can be enormous such effects are reduced significantly 
with the imposition of the EDM constraints. However, even 
with the inclusion of the EDM constraints  the effects are still very
significant and should be included in a precision prediction of 
event rates in any SUSY, string or D brane model.

\end{titlepage}

\section{Introduction}
\noindent
SUSY/string models contain soft 
parameters which are in general complex and introduce new sources
of CP violation regarding the electric dipole moment (EDM) of the electron 
and of the
neutron. The typical size of these phases in O(1) and 
they pose a serious EDM problem. Thus the current limits on
the electron\cite{commins} and the neutron\cite{harris} EDM are given by
  $|d_e|<4.3 \times 10^{-27}$ ecm, $|d_n|<6.3 \times 10^{-26}$ ecm
and an order of magnitude analysis shows that the theoretical
predictions with phases O(1) are already in excess of the experimental
limits.
For the minimal supergravity unified model (mSUGRA)\cite{chams} 
the soft SUSY breaking
sector is characterized by the parameters  
$m_0$, $m_{1/2}$, $A_0$ and $\tan\beta$,  where  $m_0$ is the 
universal scalar mass, $m_{1/2}$ is the universal gaugino mass,
$A_0$ is the universal trilinear  coupling, and $\tan\beta$ 
is defined by $\tan\beta=<H_2>/<H_1>$ where $H_2$ is the Higgs
that gives mass to the up quark and $H_1$ is  the Higgs that 
gives mass to the down quark. In addition one has the Higgs
mixing parameter $\mu$ which is viewed as  the same size as the
soft SUSY parameters, and  is determined by the constraints of
radiative breaking of the electro-weak symmetry. In mSUGRA a
set of field redefinitions shows that there are only two independent
phases in the theory, and  they can be chosen to be $\alpha_{A_0}$ 
  and $\theta_{\mu}$ where $\alpha_{A_0}$ is the phase of $A_0$ 
  and $\theta_{\mu}$ is the phase of $\mu$.

The operators that contribute to the electric dipole moments consist
of\cite{barr} 
\begin{eqnarray}
{\cal L^E}_I=-\frac{i}{2} d_f \bar{\psi} \sigma_{\mu\nu} \gamma_5 \psi
F^{\mu\nu},~~~
{\cal L^C}_I=-\frac{i}{2}\tilde d^C \bar{q} \sigma_{\mu\nu} \gamma_5 T^{a} q
 G^{\mu\nu a}\nonumber\\
{\cal L^G}_I=-\frac{1}{6}\tilde d^G f_{\alpha\beta\gamma}
G_{\alpha\mu\rho}G_{\beta\nu}^{\rho}G_{\gamma\lambda\sigma}
\epsilon^{\mu\nu\lambda\sigma}
\end{eqnarray}
Regarding the color dipole and the purely gluonic dimension six
operator one uses the so called naive dimensional analysis\cite{georgi}
$d^C_q=\frac{e}{4\pi} \tilde d^C_{q} \eta^C$,~
$d^G=\frac{eM}{4\pi} \tilde d^G \eta^G$,
where $\eta^C$ $\approx$ $\eta^G$ $\sim 3.4$ and 
 $M$ =1.19 GeV is the chiral symmetry breaking scale.
 There are several solutions suggested to control the
 EDM problem. One possibility is that the phases 
 could be small\cite{ellis,wein},
 or  there could be a mass suppression because of the largeness of the
 sparticle masses\cite{na}.  Recently, a new possibility was suggested, i.e.,
 the cancellation mechanism\cite{in1} which can control the SUSY EDM problem
 and there have  further developments\cite{in2,brhlik,accomando,in3} and
 applications\cite{pilaftsis,cin,g2,more}. 
 The cancellation mechanism works in two stages. First one typically
 has a  cancellation among the
$\tilde g, \tilde \chi^{\pm}_i, \tilde \chi_k^0$
exchange contributions to the EDMs. Second there are  further cancellation among 
the electric dipole, the chromoelectric dipole and the purely 
gluonic contributions. Such cancellations are quite generic in a
broad class of  SUSY/SUGRA\cite{in1,in2}, 
and in string and D brane models\cite{brhlik,accomando,in3}. 
In addition there
are two loop contributions involving 
axionic Higgs exchange\cite{wagner}. However, 
over most of the parameter space such contributions are relatively 
small.

While most of the analyses to explore the region of cancellations have
been numerical in nature, recently there has been an attempt to explore
 the regions of cancellations also analytically\cite{in3}. Such a  situation  
 exists in the so called scaling region\cite{scaling} where 
$\mu^2/M_Z^2>>1$ and one has 
$m_{\chi_1}\rightarrow \tilde m_1, m_{\chi_2}\rightarrow \tilde m_2$,
$m_{\chi_{3,4}}\rightarrow \mu$. 
It was shown in Ref.\cite{in3} that in the scaling 
region one cancellation point in the $m_0-m_{\frac{1}{2}}$
 plane can be promoted to a full trajectory where cancellations occur
 with only a minor adjustments of parameters. This promotion comes
 about via the following scaling on $m_0,  m_{\frac{1}{2}}$
 \begin{equation}
m_0\rightarrow \lambda m_0,
m_{\frac{1}{2}}\rightarrow \lambda m_{\frac{1}{2}}
\end{equation}
With the above scaling and 
under the constraint of the electro-weak symmetry breaking 
$\mu$ undergoes the following transformation
 $\mu \rightarrow \lambda \mu$ and the total electric dipole 
 $d_f$ transforms as 
$d_f  \rightarrow \lambda^{-2} d_f $. Thus the 
 point $d_f=0$ is invariant under  $\lambda$ scaling. Thus 
if cancellation holds at one point, it holds
at other points under scaling by only a small adjustment of parameters
and often with no adjustment of parameters at all.
As  discussed above in  mSUGRA one has  only two phases after 
field redefinitions. In the MSSM there  are  many more phases  
available\cite{in2}.
A very general analysis shows that  the electric dipole moment 
of the electron 
 $d_e$ depends on  3 phases, while the electric dipole moment of the
 neutron depends on 9 phases. Together $d_e$ and $d_n$ depend on 10 phases.
 The presence of many phases allows for cancellations in larger regions
  of the parameter space. A similar situation occurs in string and
  brane models\cite{brhlik,accomando,in3}. Of course it may happen
  that certain models turn out to be  free of the EDM problem as is the
  case in the work of Ref.\cite{babu} which also solves the
  strong CP problem. However, in general large CP phases could exist
  with a simultaneous resolution to the strong CP problem. For a 
  recent discussion of the possibilities for the resolution of the
  strong CP problem see Ref.\cite{ma}. 
   
 \section{SUSY Dark Matter}
 There are 32 new  particles in MSSM and any one of these particles 
 could be an LSP. In SUGRA models, however, one finds that  starting with 
prescribed boundary conditions at the GUT scale
 with gravity mediated breaking of supersymmetry one finds that 
 the model predicts the lightest neutralino to be the LSP over most 
 of the parameter space of the model. Further, with 
 R parity invariance the LSP will be stable and thus the 
 lightest neutralino  is predicted
 to be a candidate for cold dark matter  over most
 of the parameter space in SUGRA models. Many analyses of supersymmetric
 dark matter already exist in the literature\cite{jungman}. 
 These include effects of
 FCNC constraints from $b\rightarrow s+\gamma$\cite{bsgamma}, the effects of
 non-universalities of scalar masses\cite{nonuni,accomando1}, 
 effects of non-universalities
 of gaugino masses\cite{drees,corsetti2} and effects of 
 co-annihilation\cite{coanni}. 
 Recently, effects of
 uncertainties in the WIMP velocity in the direct and in the indirect 
 detection of dark matter have been analyzed\cite{bottino1,roskowski,corsetti}
  and analyses have  also been given
 of the effects of uncertainties of the quark mass densities on the
 direct detection rates\cite{bottino2,efo,corsetti2}. 
 In this paper we discuss the effects of 
 CP violation on direct detection. 
 
\section{CP Effects on  Dark Matter}

\begin{figure}[b]
\begin{center}
\includegraphics[width=.4\textwidth]{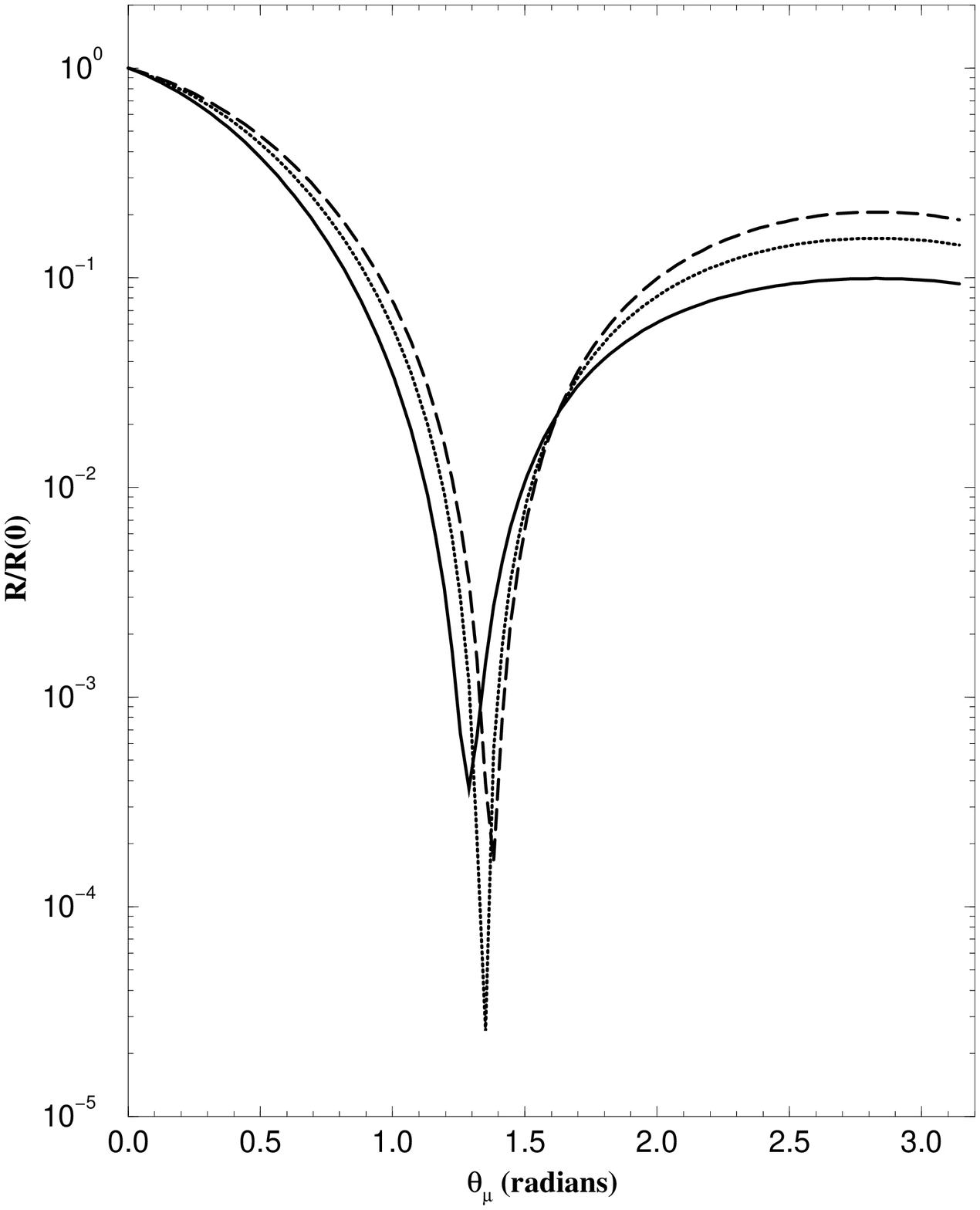}
\end{center}
\caption[]{Plot of the ratio R/R(0) vs $\theta_{\mu}$ 
without the imposition of the EDM constraints for three different 
inputs (From  Chattopadhyay et.al. in Ref.\cite{cin}). }
\label{eps1.1}
\end{figure}

    The effects of CP violation on the relic density have been
    discussed in Refs\cite{srednicki}. Here we discuss the effects of CP 
    violation on event rates\cite{ffo,cin}.   
 The effective Lagrangian with CP violation is gotten from the 
 micropscopic SUGRA lagrangian by integration on the Z, Higgs, and
 sfermion poles and one finds\cite{cin}  
 \begin{eqnarray}
 {\cal L}_{eff}=\bar{\chi}\gamma_{\mu} \gamma_5 \chi \bar{q}
\gamma^{\mu} (A P_L +B P_R) q
+ C\bar{\chi}\chi  m_q \bar{q} q+
D\bar{\chi}\gamma_5\chi  m_q \bar{q}\gamma_5 q\nonumber\\
+E  \bar{\chi}i\gamma_5\chi  m_q \bar{q}q
+F  \bar{\chi}\chi  m_q \bar{q}i\gamma_5 q
\end{eqnarray}
Here A and B are spin dependent terms arising from the
Z boson exchange and squark exhange and is given by\cite{cin}  
\begin{equation}
A=\frac{g^2}{4 M^2_W}[|X_{30}|^2-|X_{40}|^2][T_{3q}-
e_q sin^2\theta_W]
-\frac{|C_{qR}|^2}{4(M^2_{\tilde{q1}}-M^2_{\chi})}
-\frac{|C^{'}_{qR}|^2}{4(M^2_{\tilde{q2}}-M^2_{\chi})}
\end{equation}
\begin{equation}
B=-\frac{g^2}{4 M^2_W}[|X_{30}|^2-|X_{40}|^2]
e_q sin^2\theta_W +
\frac{|C_{qL}|^2}{4(M^2_{\tilde{q1}}-M^2_{\chi})}
+\frac{|C^{'}_{qL}|^2}{4(M^2_{\tilde{q2}}-M^2_{\chi})}
\end{equation}
where $C_{qR}$ etc are defined in Ref.\cite{cin} 
and $X_{n0}$ give the gaugino-Higgsino 
content of the LSP and is defined by

 \begin{equation}
 \chi^0=X_{10}^*\tilde B+X_{20}^*\tilde W+ X_{30}^*\tilde H_1+
 X_{40}^*\tilde H_2
 \end{equation}
where $\tilde B$ is the Bino, $\tilde W$  is the neutral Wino, 
and $\tilde H_1$ and $\tilde H_2$ are the Higgsinos corresponding
to the Higgs $H_1$ and $H_2$.  
In Eq.(3) C governs the scalar interaction  which arises from the 
CP even Higgs exchange and from the sfermion exhange and gives rise
to coherent scattering. It is given by\cite{cin} 
 \begin{equation}
  C=C_{\tilde{f}}+C_{h^0}+C_{H^0}
  \end{equation}
   where 
\begin{equation}
C_{\tilde{f}}(u,d)= -\frac{1}{4m_q}\frac{1}
{M^2_{\tilde{q1}}-M^2_{\chi}} Re[C_{qL}C^{*}_{qR}]
-\frac{1}{4m_q}\frac{1}
{M^2_{\tilde{q2}}-M^2_{\chi}} Re[C^{'}_{qL}C^{'*}_{qR}]
\end{equation}

\begin{equation}
C_{h^0}(u,d)=-(+)\frac{g^2}{4 M_W M^2_{h^0}}
\frac{\cos\alpha(sin\alpha)}{\sin\beta(cos\beta)} Re\sigma
\end{equation}

\begin{equation}
C_{H^0}(u,d)=
\frac{g^2}{4 M_W M^2_{H^0}}
\frac{\sin\alpha(cos\alpha)}{\sin\beta(cos\beta)} Re \rho
\end{equation}
In the above (u,d) exhibit  the quark flavor in the scattering 
and $\alpha$ stands for the  Higgs mixing angle  while   
 $\sigma$ and $\rho$ are given by
 \begin{eqnarray}
\sigma= 
 X_{40}^*(X_{20}^*-\tan\theta_W X_{10}^*)\cos\alpha
+X_{30}^*(X_{20}^*-\tan\theta_W X_{10}^*)\sin\alpha\nonumber\\
\rho=
- X_{40}^*(X_{20}^*-\tan\theta_W X_{10}^*)\sin\alpha
+X_{30}^*(X_{20}^*-\tan\theta_W X_{10}^*)\cos\alpha
\end{eqnarray}
The D term in Eq.(3) arises from the exchange of the CP odd Higgs $A^0$ 
\begin{equation}
D(u,d)= C_{\tilde{f}}(u,d)
+\frac{g^2}{4M_W}\frac{cot\beta(tan\beta)}{m_{A^0}^2}Re\omega
\end{equation}
while the terms E and F arise only in the presence of CP violation
and are given by\cite{cin}
\begin{equation}
E(u,d)=T_{\tilde{f}}(u,d)+
\frac{g^2}{4M_W} [
-(+)  \frac{cos\alpha(sin\alpha)}
{\sin\beta(cos\beta)}  \frac{Im\sigma} {m_{h^0}^2}   
+\frac{sin\alpha(cos\alpha)}
{\sin\beta(cos\beta)}\frac{Im\rho}{m_{H^0}^2}  ]
\end{equation}

\begin{figure}[b]
\begin{center}
\includegraphics[width=.4\textwidth]{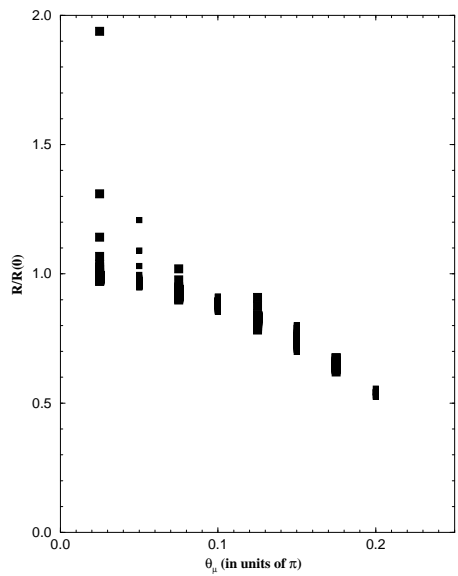}
\end{center}
\caption[]{ Scatter plot of the ratio R/R(0) vs $\theta_{\mu}$ with
inclusion of the EDM constraints (From 
Chattopadhyay et.al. in Ref.\cite{cin}). }
\label{eps1.2}
\end{figure}
   
\begin{equation}
F(u,d)=T_{\tilde{f}}(u,d)
+\frac{g^2}{4M_W} \frac{cot\beta(tan\beta)}{m_{A^0}^2}
Im\omega
\end{equation}
where $\omega$ is given by 
\begin{equation} 
 \omega= 
 -X_{40}^*(X_{20}^*-\tan\theta_W X_{10}^*)\cos\beta
+X_{30}^*(X_{20}^*-\tan\theta_W X_{10}^*)\sin\beta
\end{equation}
and 
\begin{equation}
T_{\tilde{f}}(q)= \frac{1}{4m_q}\frac{1}
{M^2_{\tilde{q1}}-M^2_{\chi}} Im[C_{qL}C^{*}_{qR}]
+\frac{1}{4m_q}\frac{1}
{M^2_{\tilde{q2}}-M^2_{\chi}} Im[C^{'}_{qL}C^{'*}_{qR}]
\end{equation}
In the limit when CP phases vanish, the above formulae limit correctly 
to previous analyses in the absence of CP phases. Numerical analysis
shows that the coefficients A-F exhibit a strong dependence on CP phases.
Typically, however, the terms D, E and F  make only small contributions
and the terms A, B and C generally dominate the scattering.
The analysis of event rates follows the method of Ref.\cite{rapn}.
The analysis including the CP violating phases but without the
imposition of the EDM constraints is displayed in Fig.1 where 
the ratio R/R(0) is plotted as a function of $\theta_{\mu}$, where
R/R(0) is the ratio of the event rates with and without CP 
violation effects.
 The analysis shows that the CP violating phases 
 can generate variations in the event rates
up to 2-3 orders of magnitude. A similar  analysis but with inclusion
of the EDM constraints in given in Fig.2. Here one finds that
 the effects are much reduced\cite{cin}, i.e., around a
 factor of 2 variation over the allowed range of phases.
 In Ref.\cite{cin} the analysis included only the two phases $\alpha_{A0}$
and $\theta_{\mu}$. However, for nonminimal models we have
three  $\xi$  phases in the the gaugino mass sector. Only one of these
three phases, i.e. $\xi_1$, enters the expressions of direct detection 
 through the neutralino mass matrix. Among the remaining two phases,
$\xi_2$ affects the EDM of the electron and of the  neutron while $\xi_3$
affects only the EDM of the neutron. Using these differential effects
generated by $\xi_1$, $\xi_2$ and $\xi_3$ we can arrange cancellations
for the EDMs to satisfy the EDM constraints and at the same time
 generate a large effect on the direct detection of neutralinos.

\section{Conclusions}
In a large class of SUSY, string and brane models there are 
 new sources of CP violation arising from the soft breaking sector
 of the theory. Since the natural size of these CP phases is 
   O(1) there exists a priori a serious EDM problem.
 The cancellation mechanism is a possible solution to the
 EDM problem with large CP phases. Detailed analyses show that 
 there exists a significant part of the parameter space where
 large CP phases are compatible with the current experiment on
 the EDMs. The existence of large CP phases can have significant
 effects on low energy SUSY phenomenology, and in this paper
 we have discussed the effects of large CP  phases on 
 event rates in the direct detection of dark matter. We emphasize that
 the inclusion of CP phases in the dark matter analysis without 
 the inclusion of EDM constraints can lead to erroneously large
 effects since the CP effects can change the event rates by several 
 orders of magnitude. With the inclusion of the EDM constraints the
 CP effects are much smaller although still significant enough to
 be included in any precision analysis of dark matter.
 These results are of import in view of the 
 ongoing\cite{dama,cdms,spooner} and
 future\cite{baudis} dark  matter experiments.   
 In addition to their effects on dark matter, 
large CP phases will also affect searches for SUSY at the
Tevatron, at the LHC and in B physics and it is imperative that
one include CP effecs in future SUSY searches to cover the 
allowed parameter space of models.
The cancellation mechanism is a testable idea. Thus if the cancellation 
idea is right, the EDMs of the electron
and of the neutron should become visible with an order of 
magnitude improvement in experiment. Such  a possibility exists
with experiments underway to improve the sensitivity of the
measurements on the electron and on the neutron electric dipole 
moments. \\

\noindent
{\bf Acknowledgements}\\
This research is supported in part by the NSF grant PHY-9901057.

%

\end{document}